\documentclass[10pt,a4paper]{article}

\usepackage{authblk}
\usepackage[utf8]{inputenc}
\usepackage[round]{natbib}
\usepackage[pdftex]{graphicx}
\usepackage{amsmath,amsthm,amsfonts}
\usepackage{lscape}   
\usepackage{rotating}
\usepackage{rotfloat}
\usepackage{multirow}
\usepackage{booktabs}
\usepackage{ctable}
\usepackage{threeparttable}
\usepackage{booktabs}
\usepackage{framed}
\usepackage{color}
\usepackage[colorlinks=true,linkcolor=blue,allcolors=blue]{hyperref}

\usepackage{soul}
\usepackage[ruled,vlined]{algorithm2e}
\usepackage{listings}
\usepackage{amsmath}
\usepackage{authblk}

\title{\texttt{FunctionalCalibration}: an R package for estimation in aggregated functional data model}
\author{Alex Rodrigo dos Santos Sousa and Vitor Ribas Perrone}
\affil{Universidade Estadual de Campinas \\ Department of Statistics, Brazil}

\date{} 

\begin{document}

\maketitle

\vspace{-1.0cm}
\begin{abstract}

  Aggregated functional data arise in a variety of applications where only linear combinations of latent functional components are observed. A prominent example is found in chemometrics through the Beer--Lambert law. This paper introduces the \texttt{FunctionalCalibration} package for \textsf{R}, which provides tools for estimating constituent curves from aggregated functional observations under additive error models. The package implements calibration methods based on B-spline and wavelet basis expansions, allowing the recovery of smooth as well as locally irregular component functions. In addition, it includes simulation tools, visualization functions, and procedures for estimating weights in prediction problems. The package is illustrated through simulated and real datasets, demonstrating its flexibility and practical applicability in functional calibration problems.
    \\

  \end{abstract}

\section{Introduction} \label{sec:intro}

The statistical problem of estimating individual curves from aggregated observations arises in several fields. In Chemometrics, for example, the Beer–Lambert law states that the absorbance curve of a substance can be represented as a weighted aggregation of the absorbance curves of its constituents, where the weights correspond to the concentrations of these constituents. The calibration problem consists of estimating the individual absorbance curves from the aggregated curve observed at discrete points. Once the model has been calibrated, it becomes possible to estimate the concentrations of the constituents in new samples, often at a substantially lower cost than performing laboratory analyses. See \cite{brereton2003} for further details on the Beer–Lambert law and calibration problems in chemometrics.

Most methods for estimating individual curves from aggregated observations are based on multivariate techniques, such as Principal Component Regression, proposed by \cite{cowe1985}, and Partial Least Squares, introduced by \cite{sjostrom1983}. Bayesian and wavelet-based approaches have also been proposed by \cite{brown1998a,brown1998b,brown2001}. More recently, methods based on Functional Data Analysis (FDA) have been developed, initially by \cite{saraiva2009}, and subsequently extended by \cite{dias-2009}, \cite{dias-2013}, \cite{dias2015}, and \cite{franco2023} using B-spline basis, and more recently by \cite{sousa-2024} using wavelet basis. 

In fact, the functional approach accounts for the intrinsic nature of the data, since the observation at discrete points is merely a limitation of the sampling process. Furthermore, the choice of basis functions depends on the characteristics of the underlying curves to be estimated. Smooth functions are generally well represented by spline basis, whereas functions exhibiting discontinuities, peaks, or oscillatory behavior are more accurately recovered using wavelet basis. See \cite{ramsay1997} and \cite{morettin2017} for background on spline and wavelet basis in FDA, respectively.

In this context, the aim of the present work is to describe the R package \texttt{FunctionalCalibration}, developed by \cite{package}, for estimating individual curves within the FDA framework. The package is publicly available on the Comprehensive R Archive Network (CRAN) (\cite{CRAN}). The package provides functions for estimating individual curves using both B-spline and wavelet basis. In particular, it implements the methods proposed by \cite{saraiva2009} and \cite{sousa-2024} for models assuming Gaussian random errors and for correlated errors under Autoregressive Fractionally Integrated Moving Average (ARFIMA) process, based on an adaptation of the method proposed by \cite{sousa-2025} in the context of nonparametric regression. 

This paper is organized as follows: Section 2 provides the statistical model and the estimation procedures under splines and wavelets basis, Sections 3 and 4 describe the functions of the package and examples respectively. Final considerations are discussed in Section 5.


\section{Theoretical background}
\subsection{Aggregated functional model}

Let $A(t)$ be a function defined as the weighted sum of $L \geq 2$ component functions given by
\begin{equation}\label{model}
A(t) = \displaystyle \sum_{l=1}^{L} y_l \alpha_l(t) + \epsilon(t),
\end{equation}
where $t \in  \mathcal{T} \subseteq \mathbb{R}$, $\alpha_l(t) \in \mathbb{L}^2(\mathbb{R}) = \{f(t):\int_\mathbb{R} f^2(t)dt < \infty\}$ are unknown functions, $y_l \in \mathbb{R}^+$ are known weights, $l = 1,\ldots,L$, and $\epsilon(t)$ is a random error term assumed to follow a Gaussian process with zero mean and unknown variance $\sigma^2$, with $\sigma > 0$. In practical terms, $A(t)$ represents the aggregated function, whereas $\alpha_l(t)$ denote the component functions underlying the aggregation. The goal here is to estimate $\alpha_l(t)$, $l=1,\ldots,L$, a task commonly referred to as the calibration problem associated with model \eqref{model}.

\medskip

In practice, the model is discretized and arises from a sample. Specifically, we suppose to observe a sample of size $N$ of aggregated functions at $M$ equally spaced points in the domain, i.e, the data consists of $\{(t_m,A_n(t_m)\}$ and $\{y_{nl}\}$ of the discretized model
\begin{equation} \label{disc_model}
A_n(t_m) = \displaystyle\sum_{l=1}^{L} y_{nl} \alpha_l(t_m) + \epsilon_n(t_m),
\end{equation}
where $n = 1,\ldots,N$ and $m = 1,\ldots,M$.

\begin{figure}[h]
\centering
\includegraphics[width=0.6\linewidth]{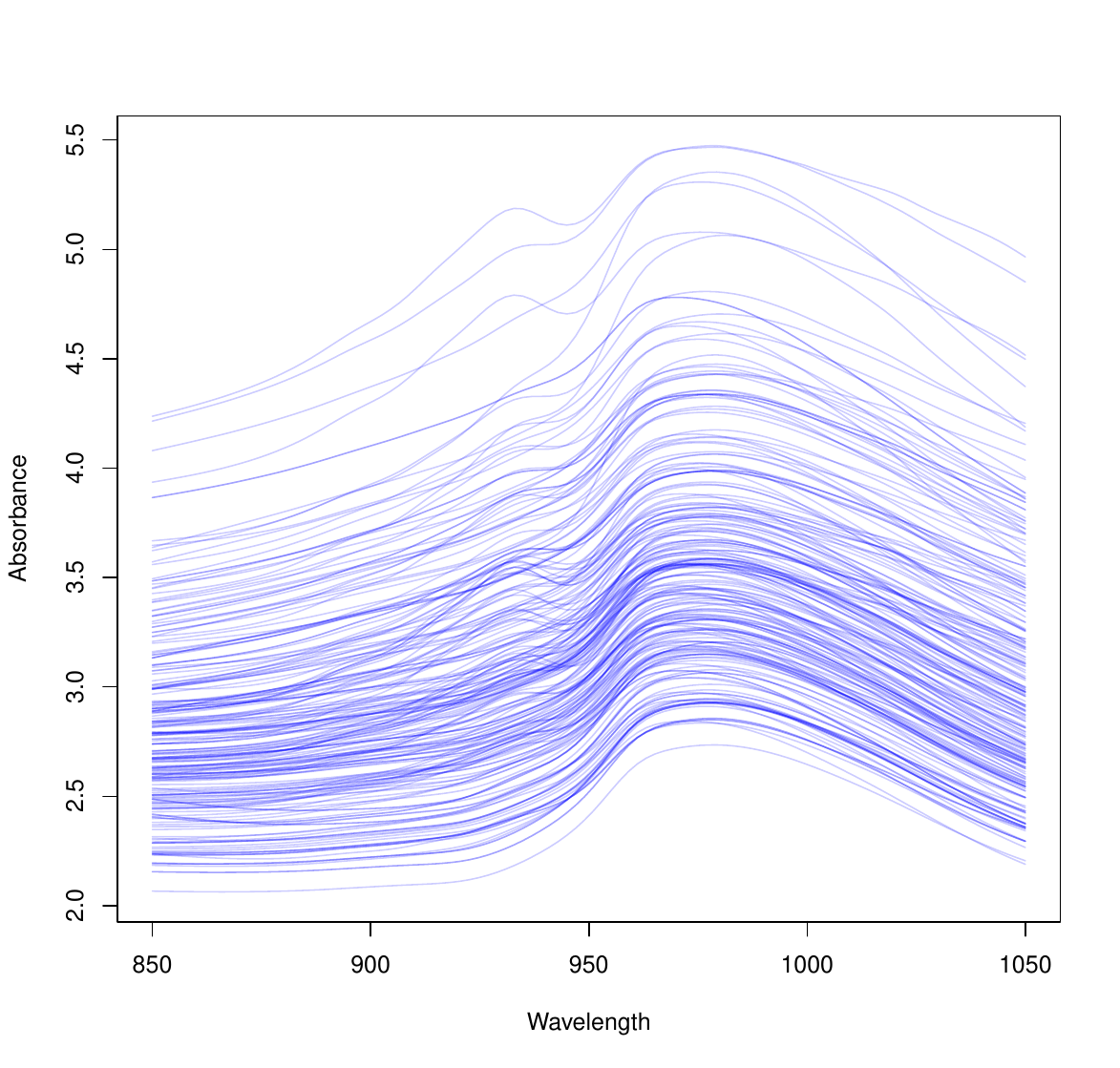}
\caption{Tecator dataset of the R package \texttt{fda.usc}.}
\label{tecator}
\end{figure}

 As an example of application of the model \eqref{model}, we have the so-called Tecator dataset, available in the \texttt{fda.usc} R package of \cite{fda_pack}. The dataset consists of $N = 215$ samples of meat absorbance curves in the Near-Infrared (NIR) spectrum measured at $M = 100$ equally spaced wavelengths, as shown in Figure \ref{tecator}. By the Beer-Lambert law (\cite{brereton2003}), the meat absorbance curve is an aggregation of the absorbance curves of protein, fat and water ($L=3$). Thus, in Chemometrics, the calibration problem consists in estimating the absorbance curves of these constituents.   

\subsection{Estimation procedures}
The standard procedure to estimate the component functions $\alpha_l(\cdot)$ of \eqref{model} is to represent them in terms of basis functions, i.e,
\begin{equation} \label{basis}
    \alpha_l(t) = \sum_{c=1}^{C} \theta_{lc}B_c(t), 
\end{equation}
for some basis functions $\{B_c(t),c=1,\ldots,C\}$ of $L^2(\mathbb{R})$. In this context, under the expansion \eqref{basis}, the problem of estimating the component functions reduces to the problem of estimating the coefficients $\theta_c$ of the basis representation. The \texttt{FunctionalCalibration} R package provides tools for estimating the component functions using wavelet and B-spline basis representations based on the proposals of \cite{sousa-2024} and \cite{saraiva2009} respectively. The next subsections present the two estimation procedures.

\subsubsection{Wavelet-based approach}

\begin{figure}[h]
\centering
\includegraphics[width=0.8\linewidth]{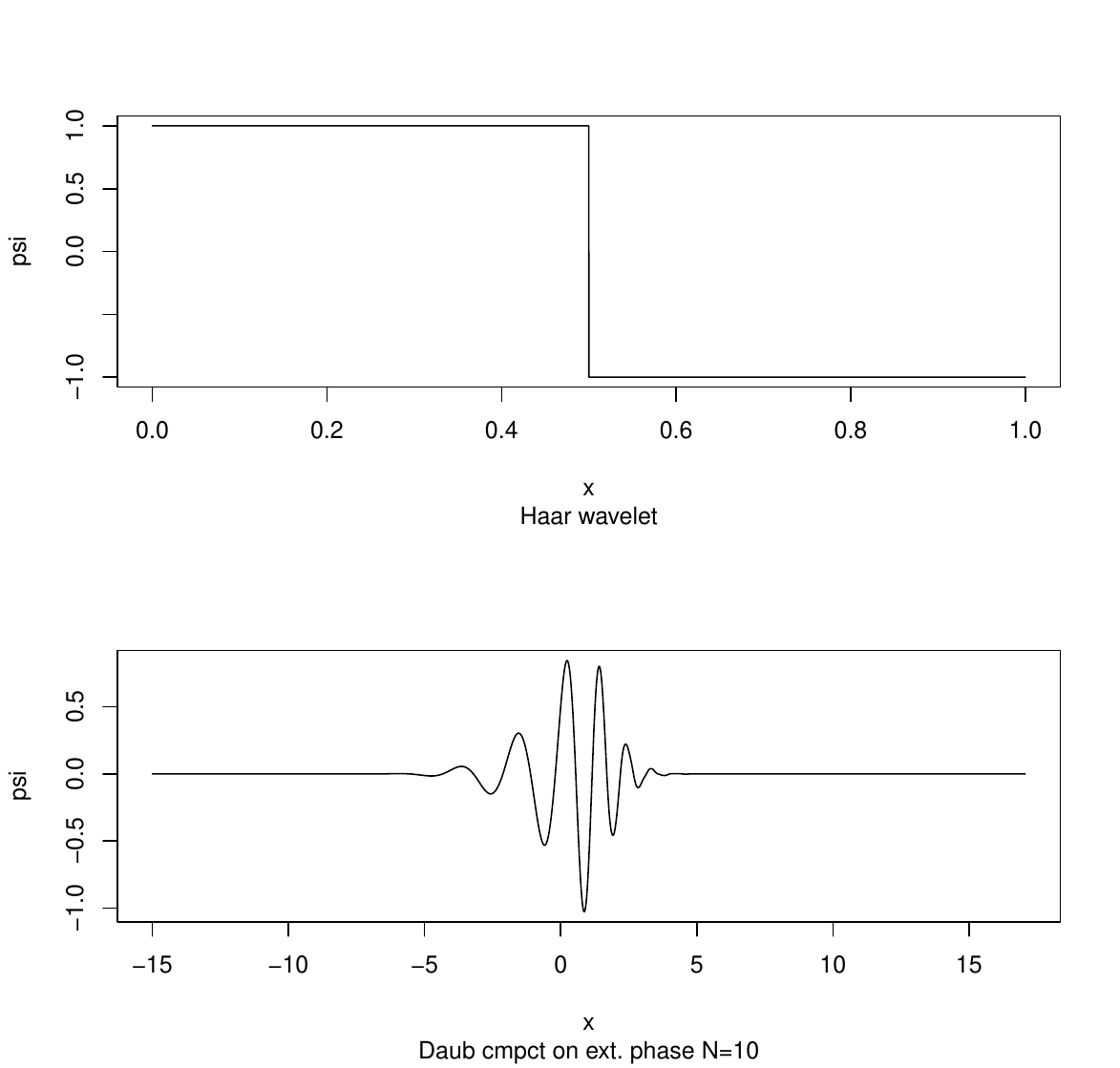}
\caption{Haar and Daubechies with 10 null moments wavelet functions.}
\label{wave_basis}
\end{figure}

\indent Under the wavelet approach, we suppose $M = 2^J$ $(J \in \mathbb{N})$. The discretized model \eqref{disc_model} can be written in matrix notation as
\begin{equation} \label{vec_model}
\mathbf{A} = \boldsymbol{\alpha y} + \boldsymbol{\epsilon},
\end{equation}
where $\boldsymbol{A} = (A_{mn}=A_n(t_m))_{1 \leq m \leq M, 1 \leq n \leq N}$, $\boldsymbol{\alpha} = (\alpha_{ml}=\alpha_l(t_m))_{1 \leq m \leq M, 1 \leq l \leq L}$, $\boldsymbol{y} = (y_{ln})_{1 \leq l \leq L, 1 \leq n \leq N}$ and $\boldsymbol{\epsilon} = (\epsilon_{mn}=\epsilon_n(t_m))_{1 \leq m \leq M, 1 \leq n \leq N}$. To perform the estimation procedure via wavelets, the functions $\alpha_l(t)$ are expanded in terms of a wavelet basis, that is, the basis representation \eqref{basis} becomes
\begin{equation} \label{wav_basis}
\alpha_l(t) = \displaystyle \sum_{j,k \in \mathbb{Z}} \theta_{ljk} , \psi_{jk}(t), \nonumber
\end{equation}
where $\{\psi_{jk}(t) = 2^{j/2}\psi(2^{j}t - k), j,k \in \mathbb{Z}\}$ is an orthonormal wavelet basis of $L^{2}(\mathbb{R})$, constructed by dilations $j$ and translations $k$ of a function $\psi$, called wavelet function or mother wavelet, and $\theta_{ljk}$ are the unknown wavelet coefficients of the expansion of the function. Figure \ref{wave_basis} shows two examples of wavelet functions: the Haar and Daubechies with 10 null moments. 

A discrete wavelet transformation (DWT), which can be represented by an orthogonal $M \times M$ transformation matrix $W$, is applied to both sides of \eqref{vec_model}, obtaining the following model in the wavelet domain,
\begin{equation} \label{wav_model}
\boldsymbol{D}= \mathbf{\Theta}\boldsymbol{y} + \boldsymbol{\varepsilon}, 
\end{equation}
where $\boldsymbol{D} = \boldsymbol{WA}$ is the matrix with the empirical (observed) wavelet coefficients,  $\boldsymbol{\Theta} = \boldsymbol{W\alpha}$ is the matrix with the wavelet coefficients to be estimated and $\boldsymbol{\varepsilon} = \boldsymbol{W\epsilon}$ is the matrix with the random errors. From this representation, in order to estimate $\mathbf{\Theta}$, a shrinkage rule $\delta(\cdot)$ is applied to each empirical coefficient $d$ of of $\boldsymbol{D}$. The estimator of $\mathbf{\Theta}$ is then obtained by
\begin{equation}
\mathbf{\hat{\Theta}} = \mathbf{\delta(D)}\boldsymbol{y}(\boldsymbol{yy}^T)^{-1}. \nonumber
\end{equation}
\indent Finally, to obtain the estimates of the functions of interest, one applies the inverse discrete wavelet transformation (IDWT), given by
\begin{equation} \label{wav_estimate}
\hat{\boldsymbol{\alpha}} = W^T\mathbf{\hat{\Gamma}}.
\end{equation}
\indent There are several shrinkage rules available in the literature, most of them based on thresholding, see \cite{vidakovic-1999}. We consider the hard- and soft-thresholding rules introduced by 
\cite{DJ-1994} and implemented in the \texttt{wavethresh} R package 
\cite{wavethresh}. These rules are respectively defined as
\begin{equation}
\delta_{\lambda}^{\mathrm{hard}}(d)
=
\begin{cases}
d, & \text{if } |d| \geq \lambda,\\
0, & \text{if } |d| < \lambda,
\end{cases}
\nonumber
\end{equation}
and
\begin{equation}
\delta_{\lambda}^{\mathrm{soft}}(d)
=
\begin{cases}
0, & \text{if } |d| \leq \lambda,\\
\operatorname{sign}(d)\left(|d|-\lambda\right),
& \text{if } |d| > \lambda,
\end{cases}
\nonumber
\end{equation}
where $\lambda>0$ denotes the threshold parameter. We consider the different
threshold-selection procedures available in the package, together with the
Bayesian shrinkage rule proposed by \cite{sousa-2022} that is based on a mixture of a point mass function at zero $\delta_0(\cdot)$ and the logistic distribution centered around zero $g(\cdot;\tau)$ as prior model,
\begin{equation} \label{prior1}
\pi(\theta; p, \tau) = p\delta_0(\theta) + (1 - p)g(\theta;\tau), 
\end{equation}
and
\begin{equation} \label{prior2}
    g(\theta; \tau) = \frac{\exp\left( -\frac{\theta}{\tau} \right)}{\tau \left( 1 + \exp\left( -\frac{\theta}{\tau} \right) \right)^2} \mathbb{I}_{\mathbb{R}}(\theta),
\end{equation}
where $p \in (0,1)$ and $\tau >0$ are the hyperparameters. The shrinkage rule associated with the prior model \eqref{prior1} and \eqref{prior2} is given by the posterior mean of $\theta$,
\begin{equation} \label{bayesrule}
\delta(d) = \mathbb{E}(\theta \mid d)= \displaystyle \frac{(1 - p) \int_{\mathbb{R}} (\hat\sigma u + d) g(\hat\sigma u + d; \tau) \phi(u) , du}{\frac{p}{\hat\sigma} \phi\left( \frac{d}{\hat\sigma} \right) + (1 - p) \int_{\mathbb{R}} g(\hat\sigma u + d; \tau) \phi(u) , du}, 
\end{equation}
\indent where $\phi(\cdot)$ denotes the standard normal probability density function.

\indent Moreover, the standard estimator of $\sigma$ and the elicitation of $p$ are given respectively by

\begin{equation}
\hat{\sigma} = \displaystyle \frac{\operatorname{median}{, |d_{J-1,k}| : k = 0, \ldots, 2^{J-1} ,}}{0.6745}, \nonumber
\end{equation}
and
\begin{equation}\label{pchoice}
    p = p(j) = 1 - \frac{1}{(j - J_0 + 1)^2}, 
\end{equation}
where $J_0$ is the primary resolution level. Moreover, \cite{sousa-2022} suggests to choose $\tau \leq10$.
See \cite{sousa-2024} for more details about the wavelet-based estimation procedure.

\subsubsection{Extension of the wavelet approach in models with correlated random error}
The wavelet-based approach described in the previous subsection can be extended to model \eqref{disc_model} to accommodate correlated random errors. In particular, we assume that the error process $\epsilon_n(t_m)$ follows an $\mathrm{AR}(1)$ structure. Due to the decorrelation property of the DWT, the shrinkage rule can be applied individually to each empirical coefficient. However, applying the rule involves estimating the standard deviation of the empirical coefficients at each resolution level, i.e, the shrinkage rule must be level-dependent, see \cite{johnstone-silerman-1997}. In this sense, we apply the Bayesian approach of \cite{sousa-2025} to the context of aggregated functional data. The shrinkage rule to be applied on a single empirical wavelet coefficient $d$ at the resolution level $j$ is
\begin{equation}\label{cor_rule}
    \delta_j(d) = \displaystyle \frac{(1 - p) \int_{\mathbb{R}} (\hat{\sigma}_j u + d) g(\hat{\sigma}_j u + d; \tau) \phi(u) du}{\frac{p}{\hat{\sigma}_j} \phi\left( \frac{d}{\hat{\sigma}_j} \right) + (1 - p) \int_{\mathbb{R}} g(\hat{\sigma}_j u + d; \tau) \phi(u) du}, \nonumber
\end{equation}
where
\begin{equation}\label{cor_sd}
    \hat{\sigma}_j = \hat{\sigma}(j) = \displaystyle \frac{\operatorname{median}\{\, |d_{j,k}| : k = 0, \ldots, 2^{j} \,\}}{0.6745}. \nonumber
\end{equation}

\subsubsection{Spline-based approach}

\begin{figure}[H]
\centering
\includegraphics[width=0.6\linewidth]{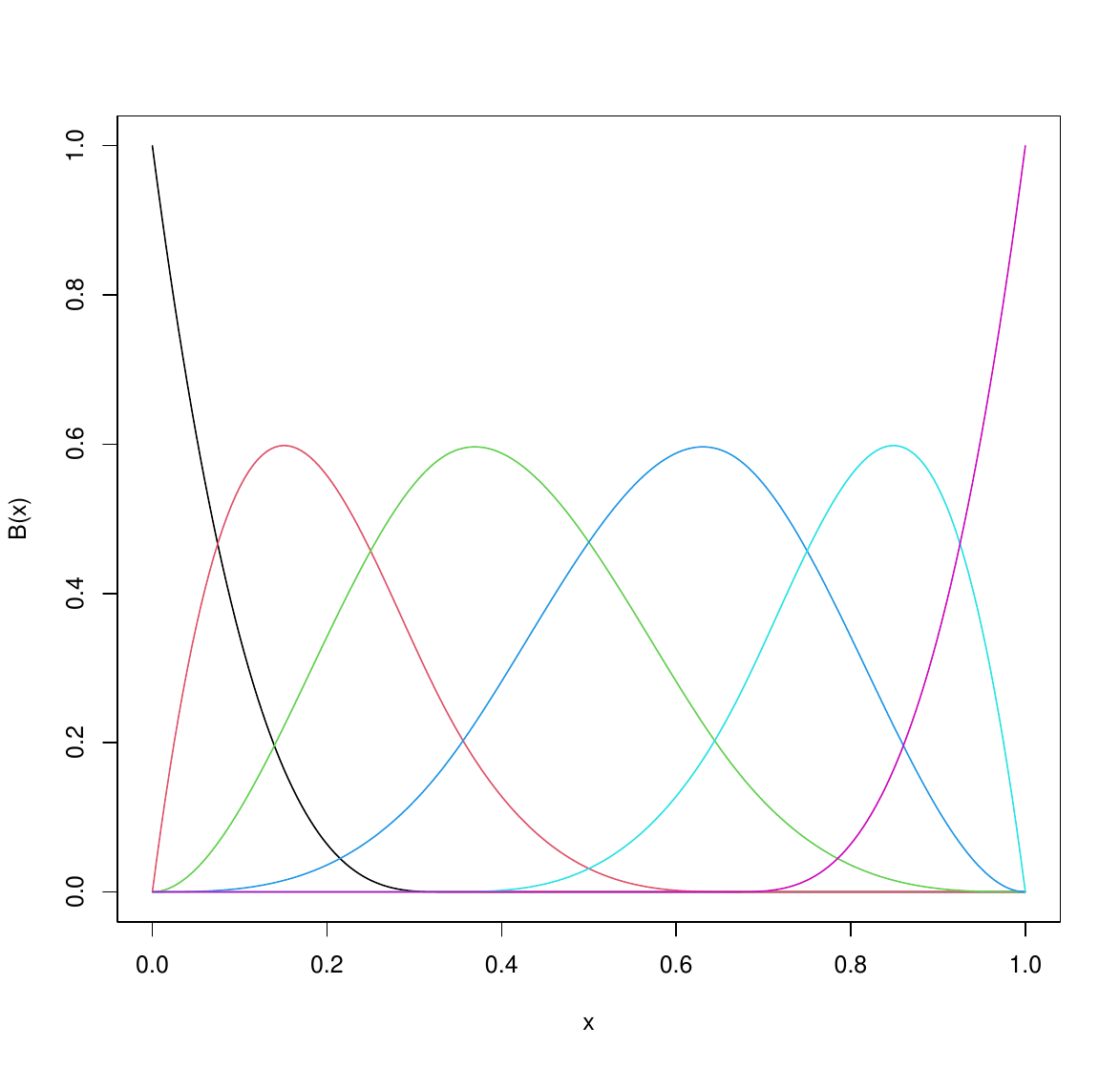}
\caption{Cubic B-splines.}
\label{spline_basis}
\end{figure}

\indent For the estimation procedure via splines, the functions $\alpha_l(t)$ of \eqref{model} are expanded in terms of B-spline basis functions, that is,

\begin{equation} \label{spline_exp}
\alpha_l(t) = \sum_{k=1}^{K}\theta_{lk} B_k(t),
\end{equation}
where $\{B_k(t)\}_{k=1}^K$ denotes a set of $K$ B-spline basis functions defined over the domain and $\theta_{lk}$ are the unknown coefficients to be estimated. Figure \ref{spline_basis} presents examples of cubic B-splines. Considering the basis expansion \eqref{spline_exp}, the discretized model \eqref{disc_model} can be written in matrix notation as
\begin{equation}
\mathbf{a} = \mathbf{B} \boldsymbol{\theta} + \boldsymbol{\epsilon}, \nonumber
\end{equation}
where $\mathbf{a}$, $\mathbf{B}$, $\boldsymbol{\theta}$ and $\boldsymbol{\epsilon}$ are given respectively by
\begin{equation}
    \begin{pmatrix}
A_1(t_1) \\
A_1(t_2) \\
\vdots \\
A_1(t_M) \\
A_2(t_1) \\
\vdots \\
A_2(t_M) \\
\vdots \\
A_N(t_1) \\
\vdots \\
A_N(t_M)
\end{pmatrix}
=
\begin{pmatrix}
y_{11}B_1(t_1) & \cdots & y_{L1}B_1(t_1) & \cdots & y_{11}B_K(t_1) & \cdots & y_{L1}B_K(t_1)\\
y_{11}B_1(t_2) & \cdots & y_{L1}B_1(t_2) & \cdots & y_{11}B_K(t_2) & \cdots & y_{L1}B_K(t_2)\\
\vdots & & \vdots & & \vdots & & \vdots\\
y_{11}B_1(t_M) & \cdots & y_{L1}B_1(t_M) & \cdots & y_{11}B_K(t_M) & \cdots & y_{L1}B_K(t_M)\\[0.2cm]

y_{12}B_1(t_1) & \cdots & y_{L2}B_1(t_1) & \cdots & y_{12}B_K(t_1) & \cdots & y_{L2}B_K(t_1)\\
\vdots & & \vdots & & \vdots & & \vdots\\
y_{12}B_1(t_M) & \cdots & y_{L2}B_1(t_M) & \cdots & y_{12}B_K(t_M) & \cdots & y_{L2}B_K(t_M)\\
\vdots & & \vdots & & \vdots & & \vdots\\
y_{1N}B_1(t_1) & \cdots & y_{LN}B_1(t_1) & \cdots & y_{1N}B_K(t_1) & \cdots & y_{LN}B_K(t_1)\\
\vdots & & \vdots & & \vdots & & \vdots\\
y_{1N}B_1(t_M) & \cdots & y_{LN}B_1(t_M) & \cdots & y_{1N}B_L(t_M) & \cdots & y_{LN}B_L(t_M)
\end{pmatrix}
\begin{pmatrix}
\theta_{01} \\
\theta_{11} \\
\vdots \\
\theta_{L1} \\
\theta_{02} \\
\theta_{12} \\
\vdots \\
\theta_{L2}\\
\vdots \\
\theta_{0K} \\
\theta_{1K} \\
\vdots \\
\theta_{LK}
\end{pmatrix}
+
\begin{pmatrix}
\epsilon_1(t_1) \\
\epsilon_1(t_2) \\
\vdots \\
\epsilon_1(t_M) \\
\epsilon_2(t_1) \\
\vdots \\
\epsilon_2(t_M) \\
\vdots \\
\epsilon_N(t_1) \\
\vdots \\
\epsilon_N(t_M)
\end{pmatrix}. \nonumber
\end{equation}

\indent The least squares estimator of $\boldsymbol{\theta}$ is 
\begin{equation} \label{spline_est}
\boldsymbol{\hat{\theta}} = (\mathbf{B}^T\mathbf{B})^{-1}\mathbf{B}^T\mathbf{a}.
\end{equation}
See \cite{saraiva2009} and \cite{dias2015} for more details about the estimation procedure and also \cite{deboor} and \cite{ramsay1997} for an overview about splines and B-splines. 

\section{Functions}

In this section we describe the seven functions of the \texttt{FunctionalCalibration} R package.

\subsection{Function: {functional\_calibration\_wavelets}}

 The function \texttt{functional\_calibration\_wavelets} estimates the component functions using a wavelet-based approach under either Gaussian independent noise (\cite{sousa-2024}) or correlated noise (\cite{sousa25}). Its implementation requires the weights $y_1,\ldots,y_L$ associated with the component functions in the linear model \eqref{model} to be known. Table \ref{tab:function_cal_wav} summarizes the function arguments and their corresponding descriptions.

\begin{table}[H]
\centering

\begin{tabular}{p{0.17\textwidth} p{0.76\textwidth}}
\hline
\textbf{Argument} & \textbf{Description} \\ \hline

\texttt{data} &
An $M \times N$ numeric matrix containing the aggregated functional data, where each column corresponds to an aggregated sample observed at $M$ domain points. \\

\texttt{weights} &
An $L \times N$ numeric matrix of weights. The $(l,i)$-th element represents the contribution of the component function $\alpha_l(x)$ to the $i$-th aggregated sample. \\

\texttt{wavelet} &
A character string specifying the wavelet family used to perform the Discrete Wavelet Transform (DWT). The default value is \texttt{"DaubExPhase"}. \\

\texttt{method} &
A character string specifying the shrinkage method applied to the empirical wavelet coefficients. Available options are \texttt{"bayesian"}, \texttt{"universal"}, \texttt{"sure"}, \texttt{"probability"}, and \texttt{"cv"}. The default value is \texttt{"bayesian"}.\\

\texttt{tau} &
A numeric value specifying the parameter $\tau$ in the Bayesian shrinkage rule. The default value is 1.\\

\texttt{p} &
A numeric value specifying the parameter $p$ in the Bayesian shrinkage rule. If not provided, it is estimated from the data. The default value is \texttt{NULL}.\\

\texttt{sigma} &
A numeric value specifying the noise standard deviation $\sigma$ in the Bayesian shrinkage rule. If not provided, it is estimated from the data. The default value is \texttt{NULL}.\\

\texttt{MC} &
A logical value indicating whether the integrals involved in the Bayesian shrinkage rule should be approximated using Monte Carlo simulation. The available values are \texttt{TRUE} and \texttt{FALSE}. The default value is \texttt{TRUE}.\\

\texttt{type} &
A character string specifying the type of thresholding rule used for non-Bayesian shrinkage methods. Available options are \texttt{"soft"} and \texttt{"hard"}. The default value is \texttt{"soft"}.\\

\texttt{singular} &
A logical value indicating whether a small constant, $1\times10^{-10}$, should be added to the diagonal of $yy^{T}$ to improve numerical stability when performing matrix inversion. The default value is \texttt{FALSE}.\\

\texttt{corre} &
A logical value indicating whether correlated errors should be considered. This option is available only for the \texttt{"bayesian"} and \texttt{"uniform"} methods. The default value is \texttt{FALSE}.\\

\texttt{x} &
A numeric vector containing the domain points at which the functional data are observed. If not provided, the default is \texttt{1:nrow(data)}. The default value is \texttt{NULL}.\\

\hline
\end{tabular}
\caption{Arguments of the function \texttt{functional\_calibration\_wavelets} and their descriptions.}
\label{tab:function_cal_wav}
\end{table}

The function returns a list with the estimated component functions $\hat{\alpha}_l(t_m)$, $l = 1,\ldots,L$ and $m = 1, \ldots,M$ and the the plots of the $L$ estimated curves. Table \ref{tab:output_cal_wav} summarizes the outputs and their respective descriptions.

\begin{table}[H]
\centering

\begin{tabular}{p{0.18\textwidth} p{0.75\textwidth}}
\hline
\textbf{Output} & \textbf{Description} \\ \hline

\texttt{alpha} &
A numeric matrix containing the estimated component functions, where the $l$-th column contains the estimated values of $\alpha_l(x)$ evaluated at the domain points. \\

\texttt{Plots} &
A list of plot objects, where the $l$-th element provides a graphical representation of the estimated component function $\alpha_l(x)$. \\

\hline
\end{tabular}
\caption{Outputs of the function \texttt{functional\_calibration\_wavelets}.}
\label{tab:output_cal_wav}
\end{table}

\subsection{Function: {functional\_calibration\_splines}}

The function \texttt{functional\_calibration\_splines} estimates the component functions of model \eqref{model} using a B-spline basis expansion, following the approach proposed by \cite{saraiva2009}. Similarly to the wavelet-based calibration function described above, this function requires the weights associated with the linear combination of the component functions to be known. Table \ref{tab:function_arguments_splines} summarizes the function arguments and their corresponding descriptions.

\begin{table}[H]
\centering
\begin{tabular}{p{0.20\textwidth} p{0.73\textwidth}}
\hline
\textbf{Argument} & \textbf{Description} \\ \hline

\texttt{data} &
An $M \times N$ numeric matrix containing the aggregated functional data, where each column corresponds to an aggregated sample observed at $M$ domain points. \\

\texttt{weights} &
An $L \times N$ numeric matrix of weights. The $(l,i)$-th element represents the contribution of the component function $\alpha_l(x)$ to the $i$-th aggregated sample. \\

\texttt{x} &
A numeric vector containing the domain points at which the functional data are observed. If not provided, the default is \texttt{1:nrow(data)}. \\

\texttt{n\_functions} &
A positive integer specifying the number of spline basis functions used to represent the component functions in the estimation procedure. The default value is 10.\\

\hline
\end{tabular}
\caption{Arguments of the function \texttt{functional\_calibration\_splines} and their descriptions.}
\label{tab:function_arguments_splines}
\end{table}

\indent As with the previous wavelet-based calibration function, \texttt{functional\_calibration\_splines} returns a list containing the estimated component functions and their corresponding plots, as summarized in Table \ref{tab:out_cal_spl}.

\begin{table}[H]
\centering

\begin{tabular}{p{0.18\textwidth} p{0.75\textwidth}}
\hline
\textbf{Output} & \textbf{Description} \\ \hline

\texttt{alpha} &
A numeric matrix containing the estimated component functions, where the $l$-th column contains the estimated values of $\alpha_l(x)$ evaluated at the domain points. \\

\texttt{plots} &
A list of plot objects, where the $l$-th element provides a graphical representation of the estimated component function $\alpha_l(x)$. \\

\hline
\end{tabular}
\caption{Outputs of the function \texttt{functional\_calibration\_splines}.}
\label{tab:out_cal_spl}
\end{table}

\subsection{Function: {plot\_aggregated\_curve}}

The function \texttt{plot\_aggregated\_curve} generates a plot of the predicted aggregated function based on the component functions estimated using either the wavelet- or spline-based approach and the corresponding known weights. More specifically, according to model \eqref{disc_model}, the predicted aggregated function for the $n$-th sample is computed as
\begin{equation} \label{pred_aggreg_curve}
\widehat{A}_n(t_m)
=
\sum_{l=1}^{L} y_{nl}\widehat{\alpha}_l(t_m),
\end{equation}
where $\widehat{\alpha}_l(t_m)$ denotes the estimated $l$-th component function evaluated at $t_m$. The function then produces a graphical representation of $\widehat{A}_n(t)$ over the observed domain. Table \ref{tab:function_argument_plot} provides the arguments of the function and their respective descriptions.
\begin{table}[H]
\centering

\begin{tabular}{p{0.18\textwidth} p{0.75\textwidth}}
\hline
\textbf{Argument} & \textbf{Description} \\ \hline

\texttt{alpha} &
A numeric matrix containing the component functions, where the $l$-th column contains the values of $\alpha_l(x)$ evaluated at the domain points specified by \texttt{x}. \\

\texttt{weights} &
A numeric vector containing the weights associated with the component functions $\alpha_l(x)$. \\

\texttt{title} &
A character string specifying the title of the plot. The default value is \texttt{NULL}.\\

\texttt{x} &
A numeric vector containing the domain points at which the component functions are evaluated. If not provided, the default is \texttt{1:nrow(alpha)}. The default value is \texttt{NULL}.\\

\hline
\end{tabular}
\caption{Arguments of the function \texttt{plot\_aggregated\_curve} and their descriptions.}
\label{tab:function_argument_plot}
\end{table}

\subsection{Function: {weight\_estimation}}

The two main functions of the package, \texttt{functional\_calibration\_wavelets} and 

\texttt{functional\_calibration\_splines}, are designed to estimate the component functions of an aggregated functional model from observed aggregated curves and a given set of weights, using either wavelet- or spline-based approaches. However, the inverse problem may also be of interest: given an observed aggregated curve and the corresponding component functions, the goal is to predict the unknown weights associated with each component. For instance, in Chemometrics, this inverse problem arises when predicting the concentrations of the constituents of a given mixture from its observed absorbance curve and the absorbance curves associated with the individual constituents, according to the Beer--Lambert law.
For this purpose, the function \texttt{weight\_estimation()} predicts the weights of the model \eqref{model} by least squares method, given by \eqref{min_sqr} in matrix notation. Table \ref{tab:function_weights} presents the arguments of the function and their descriptions.

\begin{equation}
\label{min_sqr}
    \boldsymbol{\hat{y}} = (\boldsymbol{\alpha}^T\boldsymbol{\alpha})^{-1}\boldsymbol{\alpha}^T\mathbf{A}
\end{equation}

\begin{table}[H]
\centering
\begin{tabular}{p{0.18\textwidth} p{0.75\textwidth}}
\hline
\textbf{Argument} & \textbf{Description} \\ \hline

\texttt{data} &
A numeric vector containing the observed values of the aggregated function $A(x)$ evaluated over a grid of domain points $x$. \\

\texttt{alpha} &
A numeric matrix containing the component functions, where the $l$-th column contains the values of $\alpha_l(x)$ evaluated at the same domain points as \texttt{data}. \\

\hline
\end{tabular}
\caption{Arguments of the function \texttt{weight\_estimation} and their description.}
\label{tab:function_weights}
\end{table}

\indent The function returns a numeric vector containing the weights $\hat{y}_1,\ldots,\hat{y}_L$ predicted by the least squares procedure.

The last three functions provide simulated datasets for illustrating and applying the functional calibration methods under Gaussian or correlated noise, using either wavelets or B-spline-based approaches.

\subsection{Function: {simulated\_data\_wav}}

The dataset \texttt{simulated\_data\_wav} contains $N=100$ samples of aggregated data generated from $L = 2$ component functions, $\alpha_1(x)$ and $\alpha_2(x)$, with additive independent Gaussian noise with zero mean and $\sigma^2 = 0.01$ according to model \eqref{disc_model}. The component functions used to generate the data are the Donoho and Johnstone test functions called Bumps and Doppler (\cite{DJ-1994}) given respectively by 
\begin{equation} \label{comp_fun_wav}
    \alpha_1(t) = \sum_{j=1}^{11} h_j \left(1 + \left|\frac{t - x_j}{w_j}\right|\right)^{-4}, \quad \alpha_2(t) = \sqrt{t(1-t)}\sin\left(\frac{2\pi(1 + 0.05)}{t + 0.05}\right), 
\end{equation}
where $\boldsymbol{x} = (0.1,\,0.13,\,0.15,\,0.23,\,0.25,\,0.40,\,0.44,\,0.65,\,0.76,\,0.78,\,0.81),$

$\boldsymbol{h} = (4,\,5,\,3,\,4,\,5,\,4.2,\,2.1,\,4.3,\,3.1,\,5.1,\,4.2)$ and

$\boldsymbol{w} = (0.005,\,0.005,\,0.006,\,0.01,\,0.01,\,0.03,\,0.01,\,0.01,\,0.005,\,0.008,\,0.005)$.

\begin{figure}[H]
\centering
\begin{minipage}{0.49\textwidth}
\centering
\includegraphics[width=\linewidth]{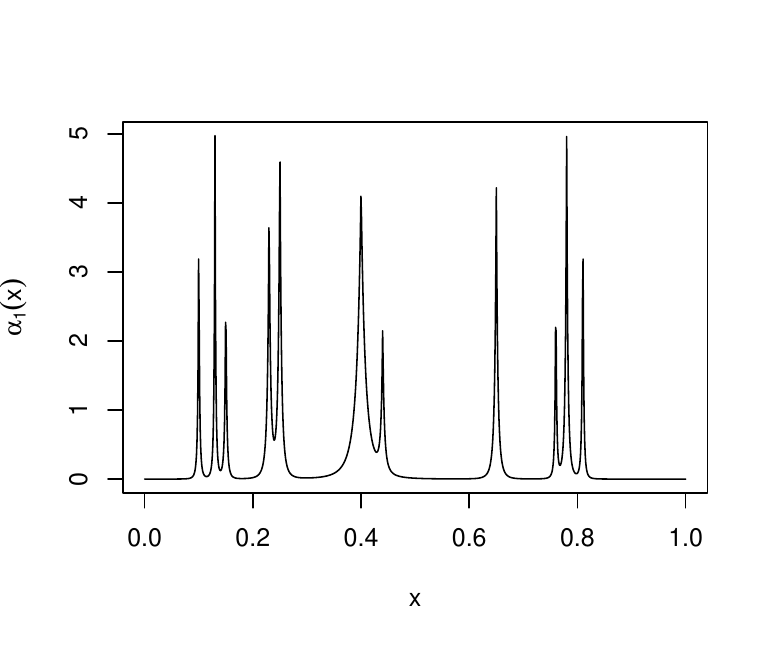}
\caption{First component function $\alpha_1(t)$ \eqref{comp_fun_wav}, called Bumps, used to simulate the dataset of {\texttt{simulated\_data\_wav}} function.}
\label{sdw_alpha1}
\end{minipage}\hfill
\begin{minipage}{0.49\textwidth}
\centering
\includegraphics[width=\linewidth]{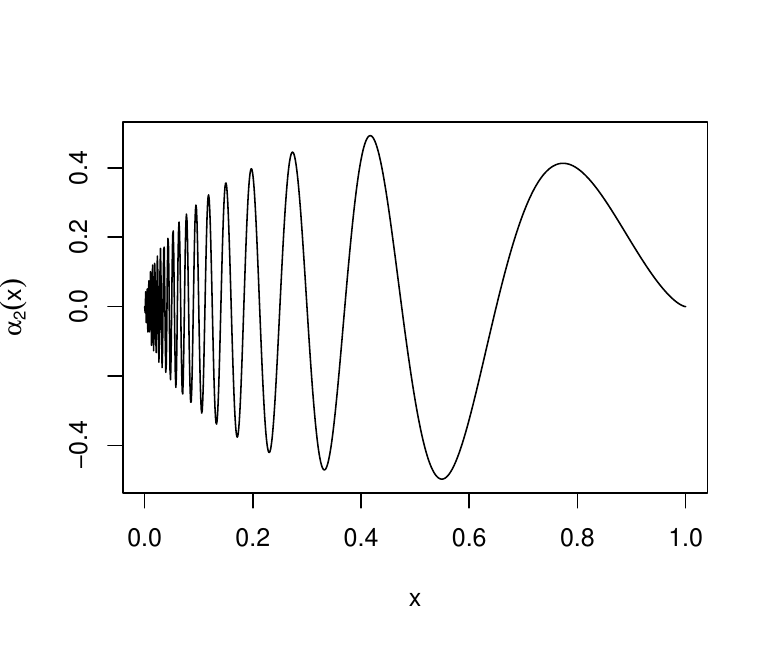}
\caption{Second component function $\alpha_2(t)$ \eqref{comp_fun_wav}, called Doppler, used to simulate the dataset of {\texttt{simulated\_data\_wav}} function.}
\label{sdw_alpha2}
\end{minipage}
\end{figure}

\begin{figure}[H]
    \centering
    \includegraphics[width=.8\linewidth]{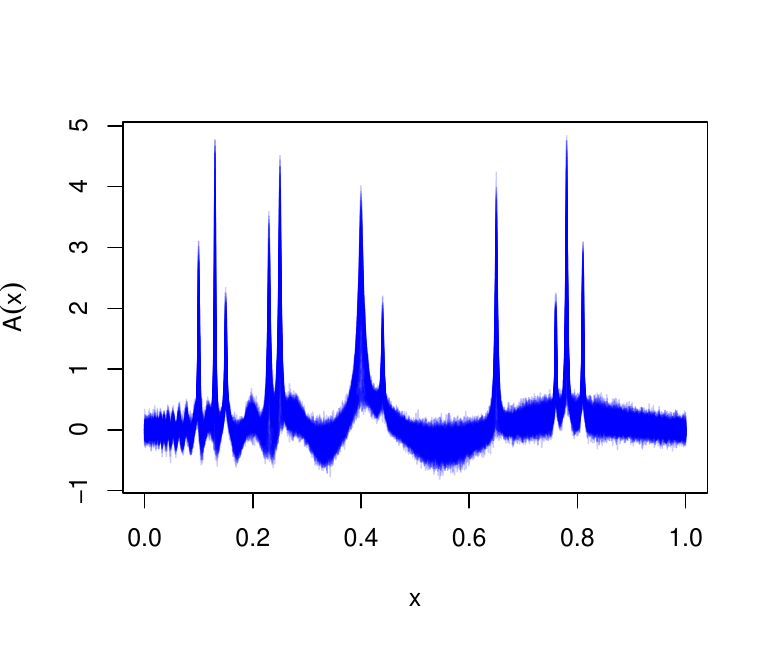}
    \caption{Dataset of aggregated curves of the \texttt{simulated\_data\_wav} function.}
    \label{agg_sdw}
\end{figure}

Figures \ref{sdw_alpha1} and \ref{sdw_alpha2} present the plots of $\alpha_1(t)$ and $\alpha_2(t)$ respectively and Figure \ref{agg_sdw} present a plot of the aggregated samples for specified weights. In fact, the component functions exhibit local features, such as spikes and oscillations, which make the wavelet-based calibration approach particularly suitable. Therefore, this dataset provides an appropriate example for illustrating the use of the \texttt{functional\_calibration\_wavelets} function under Gaussian noise.
The simulated data were carried out on an equally spaced grid of $M = 1024$ points over the interval $[0, 1]$. Table \ref{tab:wav_dataset1} shows the arguments of the function and their descriptions.

\begin{table}[H]
\centering
\begin{tabular}{p{0.18\textwidth} p{0.75\textwidth}}
\hline
\textbf{Argument} & \textbf{Description} \\ \hline

\texttt{data} &
A $1024 \times 100$ numeric matrix containing the simulated aggregated functional data with Gaussian errors. Each column corresponds to one aggregated sample observed at 1024 domain points. \\

\texttt{weights} &
A $2 \times 100$ numeric matrix containing the weights used to construct the aggregated samples from the two component functions. Each column corresponds to one sample, with weights summing to 1. \\

\texttt{x} &
A numeric vector of length 1024 containing the grid of domain points at which the functional data are observed. \\

\texttt{alphas} &
A $1024 \times 2$ numeric matrix containing the true component functions. The first and second columns contain the values of $\alpha_1(x)$ and $\alpha_2(x)$, respectively, evaluated at the 1024 domain points. \\

\hline
\end{tabular}
\caption{Arguments of the \texttt{simulated\_data\_wav} function and their descriptions.}
\label{tab:wav_dataset1}
\end{table}

\subsection{Function: {simulated\_data\_spl}}

\begin{figure}[H]
\centering
\begin{minipage}{0.49\textwidth}
\centering
\includegraphics[width=\linewidth]{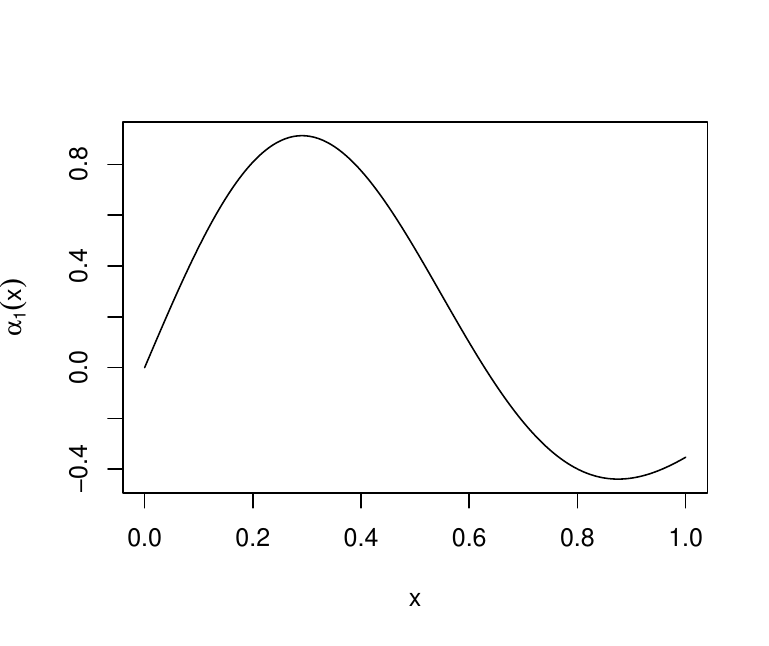}
\caption{First component function $\alpha_1(t)$ \eqref{comp_fun_spl} used to simulate the dataset of {\texttt{simulated\_data\_spl}} function.}
\label{sds_alpha1}
\end{minipage}\hfill
\begin{minipage}{0.49\textwidth}
\centering
\includegraphics[width=\linewidth]{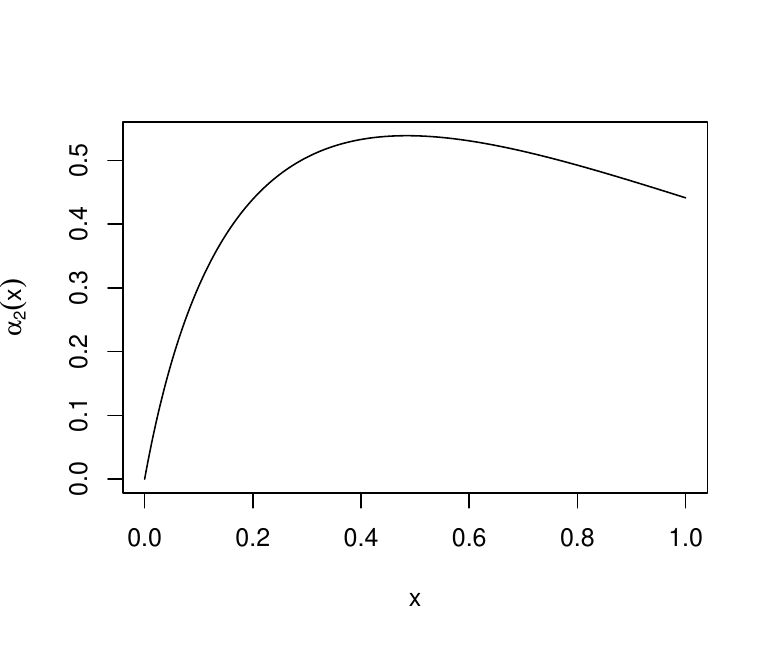}
\caption{Second component function $\alpha_2(t)$ \eqref{comp_fun_wav} used to simulate the dataset of {\texttt{simulated\_data\_spl}} function.}
\label{sds_alpha2}
\end{minipage}
\end{figure}

\begin{figure}[H]
    \centering
    \includegraphics[width=0.8\linewidth]{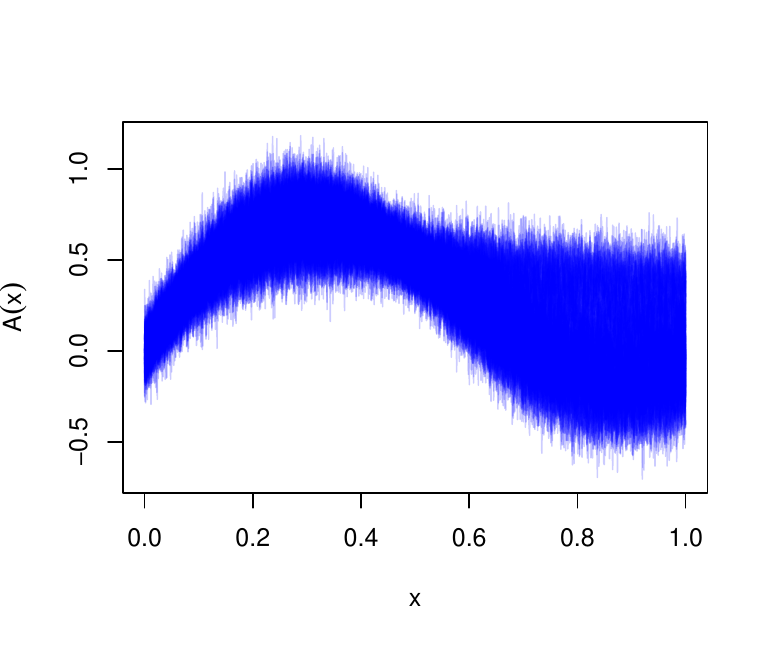}
    \caption{Dataset of aggregated curves of the \texttt{simulated\_data\_spl} function.}
    \label{agg_sds}
\end{figure}

The dataset \texttt{simulated\_data\_spl} contains $N = 100$ samples of aggregated data generated from $L=2$ component functions, $\alpha_1(t)$ and $\alpha_2(t)$, with additive independent Gaussian noise with zero mean and $\sigma^2 = 0.01$. In this case, the component functions used to generate the data are given by
\begin{equation} \label{comp_fun_spl}
    \alpha_1(t) = \sin(5t) e^{-t^2}, \quad \alpha_2(t) = \frac{1.2 \log(1 + 9t)}{\log(10)} e^{-t}.
\end{equation}
Figures \ref{sds_alpha1} and \ref{sds_alpha2} present the plots of $\alpha_1(t)$ and $\alpha_2(t)$ respectively and Figure \ref{agg_sds} present the plot of the aggregated samples for specified weights. In this case, the underlying component functions are both smooth, making the B-spline-based calibration approach particularly suitable for their estimation. Thus, this dataset provides an appropriate example for illustrating the application of the \texttt{functional\_calibration\_splines} function. The dataset were also carried out on an equally spaced grid of $M = 1024$ points over the interval $[0, 1]$. 
Table \ref{tab:spl_dataset} presents the arguments and their descriptions of the function.

\begin{table}[H]
\centering
\begin{tabular}{p{0.18\textwidth} p{0.75\textwidth}}
\hline
\textbf{Argument} & \textbf{Description} \\ \hline

\texttt{data} &
A $1024 \times 100$ numeric matrix containing the simulated aggregated functional data with Gaussian errors. Each column represents an aggregated sample observed at 1024 equally spaced domain points. \\

\texttt{weights} &
A $2 \times 100$ numeric matrix containing the weights used to construct each aggregated sample as a linear combination of the two component functions. Each column corresponds to one aggregated sample, and its elements sum to 1. \\

\texttt{x} &
A numeric vector of length 1024 containing the grid of domain points at which the aggregated and component functions are evaluated. \\

\texttt{alphas} &
A $1024 \times 2$ numeric matrix containing the true component functions. The $l$-th column contains the values of $\alpha_l(x)$ evaluated at the 1024 domain points, for $l=1,2$. \\

\hline
\end{tabular}
\caption{Arguments of the \texttt{simulated\_dataset\_spl} function and their descriptions.}
\label{tab:spl_dataset}
\end{table}

\subsection{Function: {simulated\_data\_cor}}

\begin{figure}[H]
    \centering
    \includegraphics[width=0.8\linewidth]{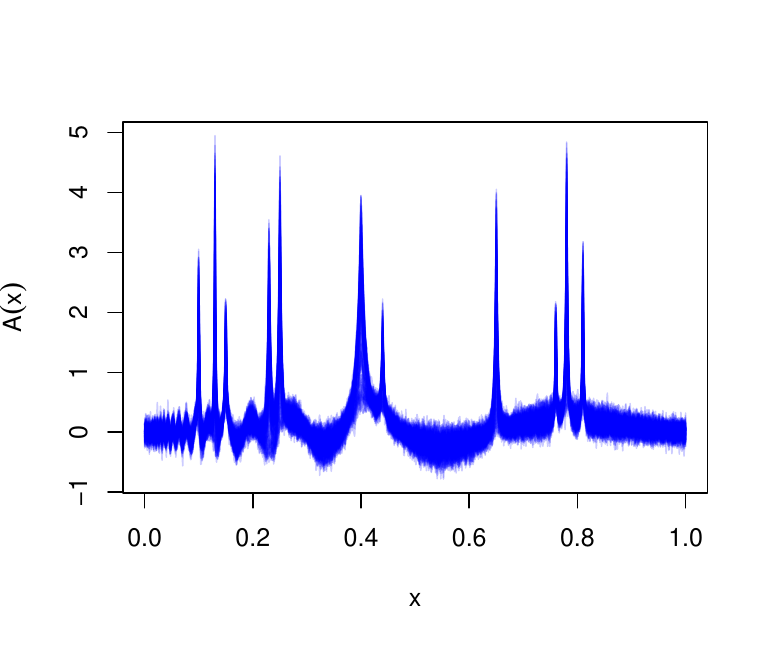}
    \caption{Dataset of aggregated curves of the \texttt{simulated\_data\_cor} function.}
    \label{agg_sdc}
\end{figure}

The last function \texttt{simulated\_data\_cor} contains $N = 100$ samples of aggregated data generated from $L=2$ component functions, $\alpha_1(t)$ and $\alpha_2(t)$, with additive correlated noise following an $\mathrm{AR}(1)$ process with autoregressive parameter $\phi = 0.5$. The underlying component functions are the Bumps and Doppler functions \eqref{comp_fun_wav} used in the data generation of \texttt{simulated\_data\_wav}. Consequently, this dataset can be used to demonstrate the performance of the wavelet-based calibration procedure in the presence of correlated noise through the function \texttt{simulated\_data\_wav}. Figure \ref{agg_sdc} present the plot of the aggregated samples for specified weights. As in the previous simulated datasets, the simulations were carried out on an equally spaced grid of $M = 1024$ points over the interval $[0, 1]$. The arguments and descriptions of the function are presented in Table \ref{tab:simulated_dataset_ar1}.

\begin{table}[H]
\centering

\begin{tabular}{p{0.18\textwidth} p{0.75\textwidth}}
\hline
\textbf{Argument} & \textbf{Description} \\ \hline

\texttt{data} &
A $1024 \times 100$ numeric matrix containing the simulated aggregated functional data with AR(1) errors. Each column represents an aggregated sample observed at 1024 equally spaced domain points. \\

\texttt{weights} &
A $2 \times 100$ numeric matrix containing the weights used to construct each aggregated sample as a linear combination of the two component functions. Each column corresponds to one aggregated sample, and its elements sum to 1. \\

\texttt{x} &
A numeric vector of length 1024 containing the grid of domain points at which the aggregated and component functions are evaluated. \\

\texttt{alphas} &
A $1024 \times 2$ numeric matrix containing the true component functions. The $l$-th column contains the values of $\alpha_l(x)$ evaluated at the 1024 domain points, for $l=1,2$. \\

\hline
\end{tabular}
\caption{Arguments of the \texttt{simulated\_data\_cor} function and their descriptions.}
\label{tab:simulated_dataset_ar1}
\end{table}

\section{Examples}
In this section, we illustrate the use of the proposed functions through practical examples. To this end, the calibration methods are applied to the simulated datasets available in the package.

\subsection{Wavelet-based functional calibration under Gaussian noise}

Let us consider the dataset available through the \texttt{simulated\_data\_wav} function. The dataset consists of $N=100$ samples of aggregated curves, each evaluated at $M=1024$ points, together with the corresponding weights of the $L=2$ underlying component functions. These functions are estimated through a wavelet expansion using the \texttt{functional\_calibration\_wavelets} function, with Daubechies wavelets with ten vanishing moments (Daub10) and the Bayesian shrinkage rule given in \eqref{bayesrule}, setting $\tau=1$ and using the automatic selection of $p$ defined in \eqref{pchoice}. The corresponding R code and its outputs are given by
\begin{verbatim}
R> functional_calibration_wavelets(simulated_data_wav$data,
+ simulated_data_wav$weights, wavelet = "DaubExPhase", method = "bayesian", tau = 1)

$alpha
                 [,1]          [,2]
   [1,] -3.446089e-02 -0.0052696898
   [2,] -2.764083e-02 -0.0081289975
   [3,] -2.726405e-02 -0.0071128458
   [4,] -2.416447e-02 -0.0082892814
   [5,] -2.156771e-02 -0.0091050315
   [ reached getOption("max.print") -- omitted 1019 rows ]
$Plots
$Plots[[1]]

$Plots[[2]]
\end{verbatim}

\begin{figure}[H]
\centering
\begin{minipage}{0.49\textwidth}
\centering
\includegraphics[width=\linewidth]{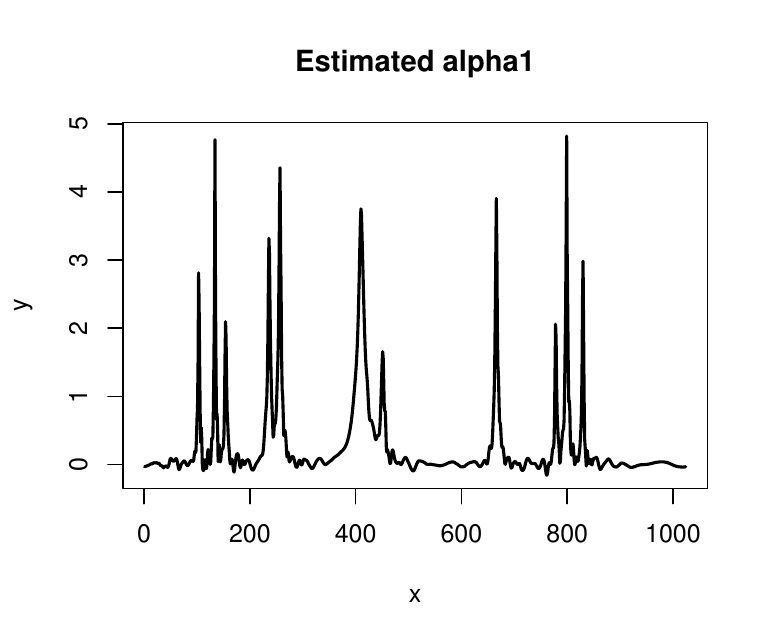}
\caption{Estimated first component function $\hat{\alpha_1}(t)$ of the \texttt{simulated\_data\_wav} dataset.}
\label{wav_est_alpha1}
\end{minipage}\hfill
\begin{minipage}{0.49\textwidth}
\centering
\includegraphics[width=\linewidth]{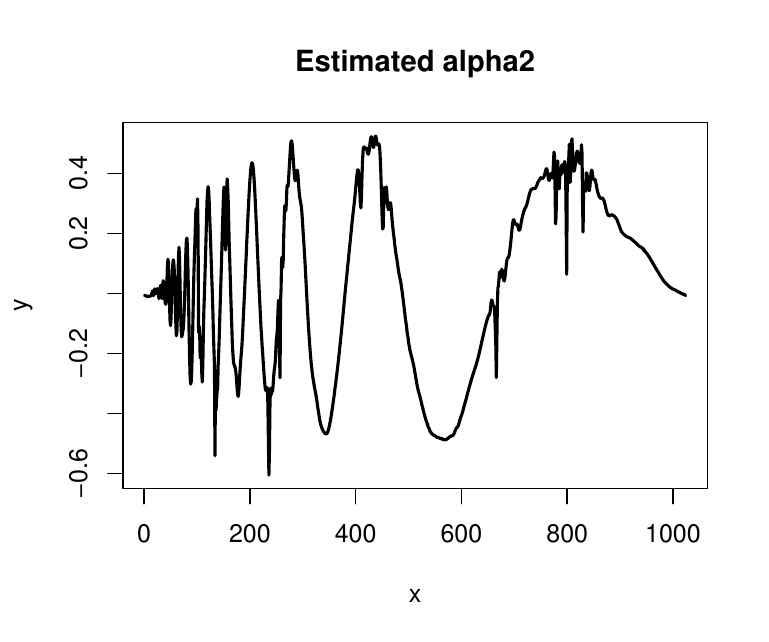}
\caption{Estimated second component function $\hat{\alpha}_2(t)$ of the \texttt{simulated\_data\_wav} dataset.}
\label{wav_est_alpha2}
\end{minipage}
\end{figure}

Figures \ref{wav_est_alpha1} and \ref{wav_est_alpha2} present the estimated curves $\hat{\alpha}_1(t)$ and $\hat{\alpha}_2(t)$ respectively. Indeed, the estimated curves successfully capture the main local features and oscillatory behavior of the underlying Bumps and Doppler component functions.

\subsection{Wavelet-based functional calibration under correlated noise}

Since the \texttt{functional\_calibration\_wavelets} function can also handle aggregated data with correlated noise, we consider the dataset available through the \texttt{simulated\_data\_cor} function to illustrate the calibration procedure under this setting. In this case, the argument \texttt{corre = TRUE} is specified when calling the function:

\begin{verbatim}
R> functional_calibration_wavelets(simulated_data_cor$data, 
+ simulated_data_cor$weights, wavelet = "DaubExPhase",
+ method = "bayesian", tau = 1, corre = TRUE)

$alpha
                 [,1]          [,2]
   [1,]  6.941096e-02  0.1094827892
   [2,]  7.356505e-02  0.1102622029
   [3,]  7.256784e-02  0.1219073869
   [4,]  7.294315e-02  0.1028052794
   [5,]  7.292351e-02  0.1103708174
   [ reached getOption("max.print") -- omitted 1019 rows ]
$Plots
$Plots[[1]]

$Plots[[2]]
\end{verbatim}

\begin{figure}[H]
\centering
\begin{minipage}{0.49\textwidth}
\centering
\includegraphics[width=\linewidth]{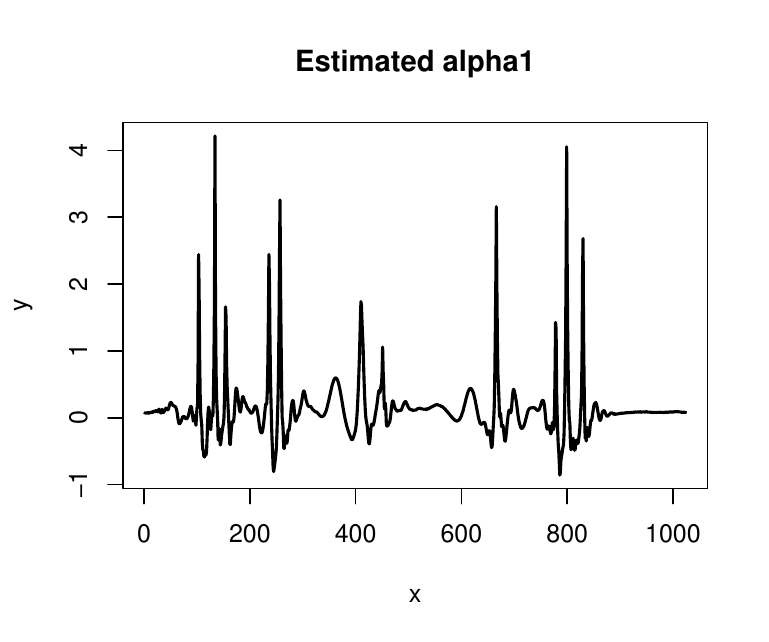}
\caption{Estimated first component function $\alpha_1(x)$ of the \texttt{simulated\_data\_cor} dataset.}
\label{cor_est_alpha1}
\end{minipage}\hfill
\begin{minipage}{0.49\textwidth}
\centering
\includegraphics[width=\linewidth]{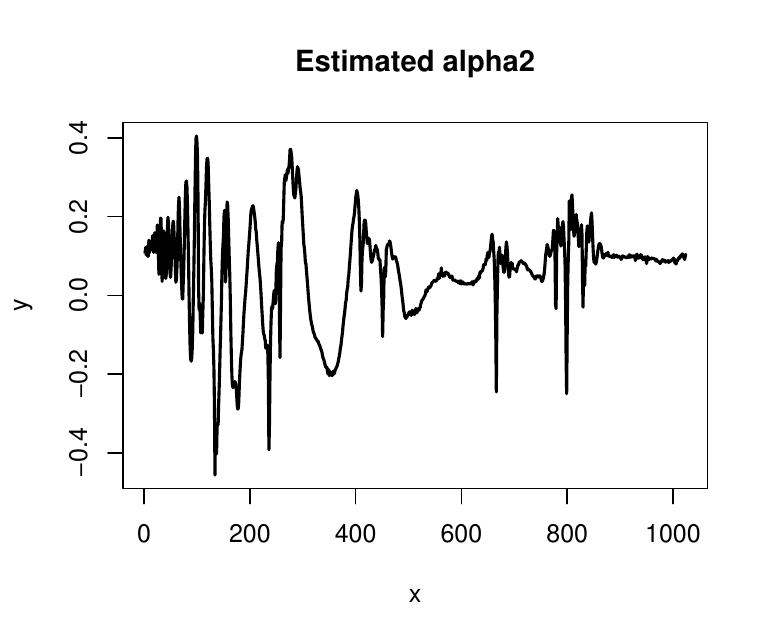}
\caption{Estimated second component function $\alpha_2(x)$ of the \texttt{simulated\_data\_cor} dataset.}
\label{cor_est_alpha2}
\end{minipage}
\end{figure}

 Figures \ref{cor_est_alpha1} and \ref{cor_est_alpha2} provide the estimated curves $\hat{\alpha}_1(t)$ and $\hat{\alpha}_2(t)$ respectively.

\subsection{Splines-based functional calibration}

In order to illustrate the functional calibration by B-splines expansion, we now consider the dataset of the \texttt{simulated\_data\_spl} function, i.e, $N = 100$ samples of aggregated curves evaluated at $M = 1024$ points. The following code estimated the $L = 2$ underlying component functions using ten cubic B-splines basis in the \texttt{functional\_calibration\_splines} function:
\begin{verbatim}
R> functional_calibration_splines(simulated_data_spl$data,
+ simulated_data_spl$weights, simulated_data_spl$x, n_functions = 10)

$alpha
                [,1]        [,2]
   [1,]  0.000000000 0.000000000
   [2,]  0.005013702 0.004260079
   [3,]  0.010024452 0.008489584
   [4,]  0.015032136 0.012688665
   [5,]  0.020036641 0.016857473
   [ reached getOption("max.print") -- omitted 1019 rows ]
$Plots
$Plots[[1]]

$Plots[[2]]
 \end{verbatim}

\begin{figure}[H]
\centering
\begin{minipage}{0.49\textwidth}
\centering
\includegraphics[width=\linewidth]{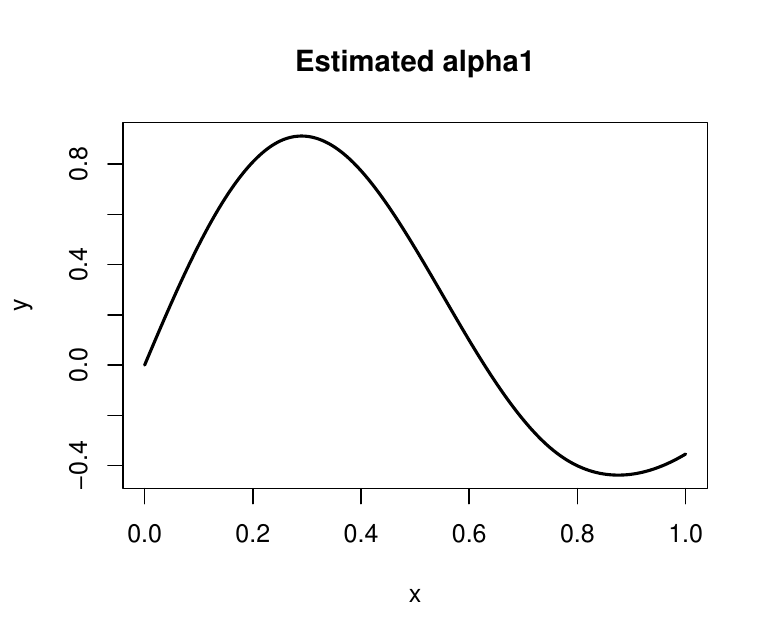}
\caption{Estimated first component function $\hat{\alpha}_1(t)$ of the \texttt{simulated\_data\_spl} dataset.}
\label{spl_est_alpha1}
\end{minipage}\hfill
\begin{minipage}{0.49\textwidth}
\centering
\includegraphics[width=\linewidth]{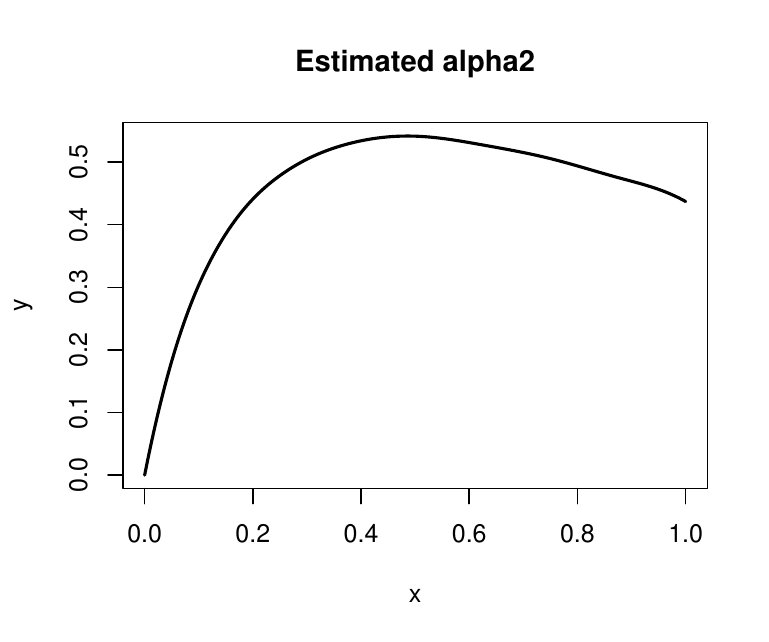}
\caption{Estimated second component function $\hat{\alpha}_2(t)$ of the \texttt{simulated\_data\_spl} dataset.}
\label{spl_est_alpha2}
\end{minipage}
\end{figure}

Figures \ref{spl_est_alpha1} and \ref{spl_est_alpha2} present the estimated curves $\hat{\alpha}_1(t)$ and $\hat{\alpha}_2(t)$ respectively. Both the estimated curves recovered the smooth feature of the true component functions. 

\subsection{Predicting the weights and plotting the aggregated function}

Having illustrated the main functions directly related to the calibration procedure, we now introduce the supporting functions available in the package. In particular, for the prediction problem, where the goal is to estimate the weights from the observed aggregated curve and the corresponding component functions, the function \texttt{weight\_estimation} provides a useful tool for obtaining such estimates. Thus, let us consider the first column of the dataset provided by the \texttt{simulated\_data\_wav} function, corresponding to the first sample of the aggregated curve $A(t)$, namely, $A_1(t_1),\ldots,A_1(t_{1024})$, which can be accessed through \texttt{simulated\_data\_wav\$data[,1]}. We also consider the corresponding known component functions $\alpha_1(t)$ and $\alpha_2(t)$, evaluated at $M=1024$ equally spaced points on $[0,1]$ and available through \texttt{simulated\_data\_wav\$alphas}. The associated weights $y_{11}$ and $y_{12}$ in model \eqref{disc_model} can then be estimated as follows:

\begin{verbatim}
R> weight_estimation(simulated_data_wav$data[,1], simulated_data_wav$alphas)

0.1014095 0.8949011 
\end{verbatim}

\indent Since the dataset is simulated, the true parameter values are known. Therefore, the estimated weights can be directly compared with their corresponding true values. These values can be obtained from the first column of the weight matrix as follows:

\begin{verbatim}
R> simulated_data_wav$weights[,1]

0.09702885 0.90297115
\end{verbatim}

Furthermore, given a set of component functions, any linear combination of them can be readily visualized using the \texttt{plot\_aggregated\_curve} function. For instance, applying this function to the component functions from the \texttt{simulated\_data\_wav} dataset with the weight vector $(0.7,0.3)$ yields the aggregated curve of Figure \ref{plot_agg_curve}, obtained according to \eqref{pred_aggreg_curve}:
\begin{verbatim}
R> plot_aggregated_curve(simulated_data_wav$alphas, c(0.7, 0.3), 
+ "Aggregated Curve Example", simulated_data_wav$x)
\end{verbatim}

\begin{figure}
    \centering
    \includegraphics[width=0.5\linewidth]{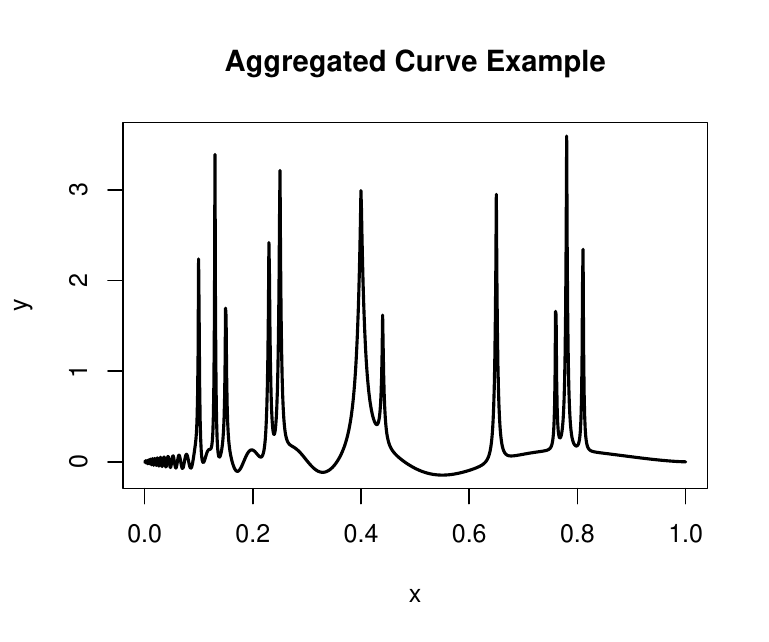}
    \caption{Aggregated curve example using component functions from \texttt{simulated\_data\_wav} dataset with weight vector $(0.7, 0.3)$.}
    \label{plot_agg_curve}
\end{figure}

\subsection{Tecator dataset}
Finally, we illustrate the package to calibrate the model \eqref{disc_model} for the Tecator dataset described in Section 2 and plotted in Figure \ref{tecator}. According to the Beer-Lambert law, the meat absorbance curve $A_{meat}(t)$ can be represented in terms of the absorbance curves of fat ($\alpha_{fat}(t)$), protein ($\alpha_{protein}(t)$) and water ($\alpha_{water}(t)$) as
\begin{equation} 
    A_{meat}(t) = y_{fat} \alpha_{fat}(t) + y_{protein} \alpha_{protein}(t) + y_{water} \alpha_{water}(t) + e(t), \nonumber
\end{equation}
where $y_{fat}$, $y_{protein}$ and $y_{water}$ are the their respective concentrations.

Since the absorbance curves are smooth, the calibration via B-spline is suitable. The following code does the calibration in order to estimate the absorbance curves of fat, protein and water using ten cubic B-splines basis:

\begin{verbatim}
library(fda.usc)

# 1. Data: 215 spectra observed at 100 wavelengths
data("tecator", package = "fda.usc")

# The package requires an M x N data matrix: evaluation points in the rows
# and samples in the columns.
absorbance <- t(as.matrix(tecator$absorp.fdata$data))
wavelength <- as.numeric(tecator$absorp.fdata$argvals)

composition <- as.data.frame(tecator$y)
variable_names <- tolower(names(composition))

find_column <- function(variable) {
  index <- match(variable, variable_names)
  if (is.na(index)) {
    stop(sprintf(
      "Column '%s' was not found. Available columns: %s",
      variable, paste(names(composition), collapse = ", ")
    ))
  }
  index
}

# L x N weight matrix, ordered as water, protein, and fat.
weights <- rbind(
  Water = as.numeric(composition[[find_column("water")]]),
  Protein = as.numeric(composition[[find_column("protein")]]),
  Fat = as.numeric(composition[[find_column("fat")]])
)

stopifnot(
  nrow(absorbance) == length(wavelength),
  ncol(absorbance) == ncol(weights),
  qr(t(weights))$rank == nrow(weights)
)

par(mfrow = c(2, 2))
# 2. Functional calibration using B-splines
# Tecator has M = 100, so the spline approach can be applied directly.
fit <- functional_calibration_splines(
  data = absorbance,
  weights = weights,
  x = wavelength,
  n_functions = 10
)
curves <- fit$alpha
\end{verbatim}

\begin{figure}[h]
    \centering
    \includegraphics[width=1\linewidth]{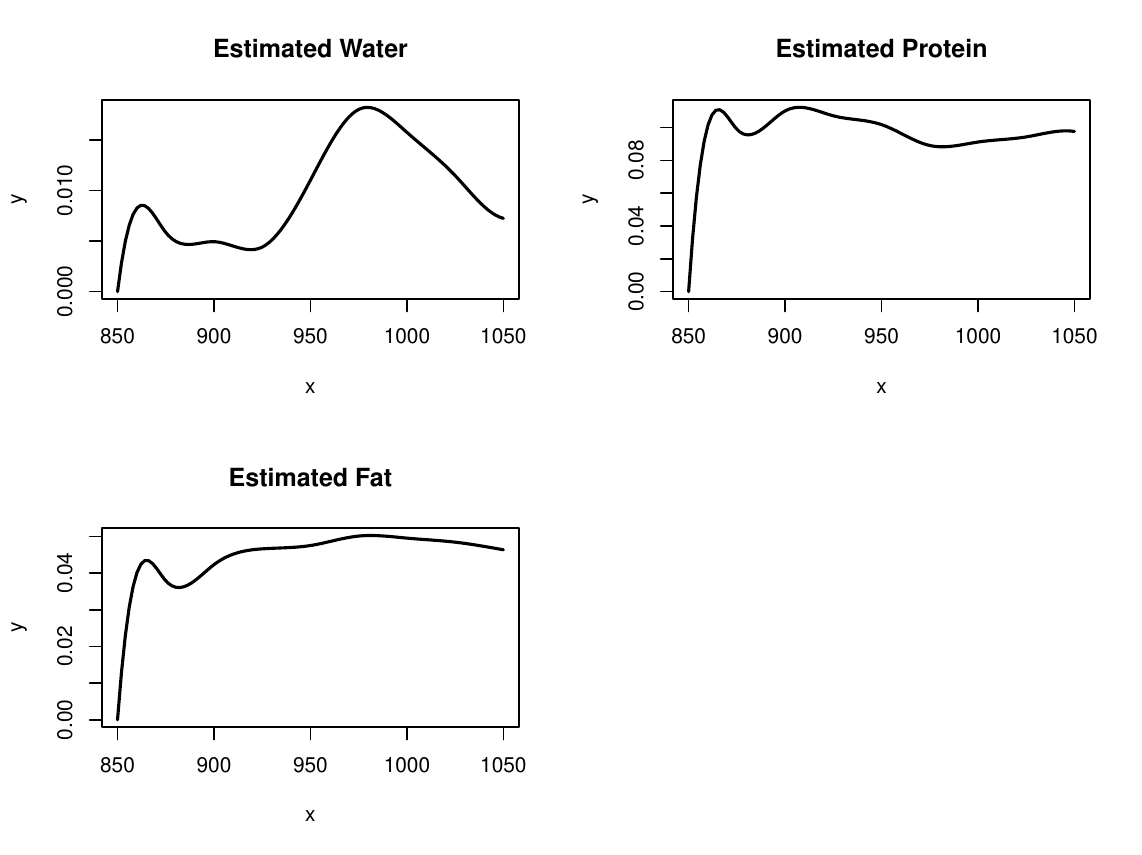}
    \caption{Estimated absorbance curves of water, protein and fat of the Tecator dataset by spline-based calibration.}
    \label{plot_tecator_abs}
\end{figure}

Figure \ref{plot_tecator_abs} presents the estimated absorbance curves $\hat{\alpha}_{fat}(t)$, $\hat{\alpha}_{protein}(t)$ and $\hat{\alpha}_{water}(t)$ of fat, protein and water respectively.

\section{Summary and discussion} \label{sec:summary}

The \texttt{FunctionalCalibration} package provides a computational framework for the analysis of aggregated functional data, with particular emphasis on the recovery of latent component functions from observed aggregated curves. The package implements two complementary estimation strategies based on B-spline and wavelet representations, allowing the user to select the approach that is more appropriate for the underlying structure of the component functions.

The spline-based procedure is especially suitable for smooth component functions, whereas the wavelet-based approach offers greater flexibility for functions exhibiting local features, abrupt changes, spikes, or oscillatory behavior. In addition, the wavelet methodology can accommodate both independent Gaussian errors and correlated errors, increasing the applicability of the package to more complex functional data settings.

Besides estimating the component functions when the aggregation weights are known, the package also addresses the inverse problem of estimating the weights from an observed aggregated curve and a given set of component functions. This feature broadens the scope of the package and makes it particularly useful in applications such as chemometrics, where the weights may represent concentrations of different constituents in a mixture.

The package also includes graphical tools and simulated datasets designed to facilitate the implementation, interpretation, and comparison of the available methods. Overall, \texttt{FunctionalCalibration} provides an integrated and accessible environment for functional calibration problems and may serve as a useful computational tool for both methodological research and practical applications.


\section*{Computational details}

The results in this paper were obtained using
\texttt{R}~3.2.4 with the
\texttt{wavethresh}~4.7.3 package. \texttt{R} itself
and all packages used are available from the Comprehensive
\texttt{R} Archive Network (CRAN) at
\url{https://CRAN.R-project.org/}.

\section*{Acknowledgments}
The authors were financed by the Programa de Incentivo a
Novos Docentes (PIND/FAEPEX) of Universidade Estadual de Campinas, grant 3376/23.

\bibliographystyle{plainnat}
\bibliography{references}



\end{document}